\documentstyle[12pt]{article}
\begin{document}
\begin{center}
\null\vskip-1truecm
\rightline{IC/94/396}
\vskip1truecm International Atomic Energy Agency\\
and\\
United Nations Educational Scientific and Cultural
Organization\\
\medskip INTERNATIONAL CENTRE FOR THEORETICAL PHYSICS\\
\vskip2truecm
{\bf ON THE OVERLAP FORMULATION\\ OF CHIRAL GAUGE THEORY}
\vskip1.2truecm
S. Randjbar-Daemi \quad and\quad J. Strathdee\\
 International Centre
for Theoretical Physics, Trieste, Italy.\\
\end{center}
\vskip1.5truecm

\centerline{ABSTRACT}
\baselineskip=20pt
\bigskip

The overlap formula proposed by Narayanan and Neuberger in chiral
gauge theories is examined. The free chiral and Dirac Green's
functions are constructed in this formalism. Four dimensional
anomalies are calculated and the usual anomaly cancellation for
one standard family of quarks and leptons is verified.
\vskip5truecm

\begin{center}
 MIRAMARE -- TRIESTE\\
\medskip
December 1994
\end{center}

\newpage

\baselineskip=20pt
\section{Introduction}

Non--perturbative study of the standard electroweak model has long
been  obstructed by difficulties with lattice regularization of
theories  involving chiral fermions. Recently, however, there has
been some progress in meeting these difficulties. In particular,
following an earlier work of Kaplan [1], an interesting new
interpretation of the chirality problem has been proposed by
Narayanan and Neuberger [2]. These authors claim that their
approach circumvents the well known no--go theorems [3] and thus has
the potential for application to numerical study of chiral symmetry
breaking and other important problems. Their basic idea is to
represent the vacuum amplitude for a chiral fermion as the overlap
of two well--defined vectors that can be interpreted as ground
states of massive fermions in an auxiliary 5--dimensional
spacetime. They show that this overlap transforms anomalously, as
it should. On the other hand, it can be regularized by any of the
standard methods, including the usual lattice regularization for
massive, non--chiral fermions. This reconciliation of features long
thought to be incompatible is a remarkable achievement.

Our purpose in this note is to explore this mechanism by
constructing the free fermion Green's function in the auxiliary
spacetime to see how the chiral Green's function emerges when the
auxiliary mass is taken to infinity. We also verify that the
4--dimensional vacuum amplitude expresses the expected perturbative
anomalies.
Finally, we show how a single anomaly free family of leptons and
quarks can be represented in this scheme.

\section{Green's functions for free chiral fermions}

The idea is that the Green's functions for a quantum field theory
in 4--dimensional Euclidean spacetime can be given an operator
realization in the Hilbert space associated with a 4+1--dimensional
(Minkowskian) spacetime. This Hilbert space contains many states
that are not relevant to the 4--dimensional physics and these are
projected out by taking the limit $\vert\Lambda\vert\to\infty$,
where $\Lambda$ is a ``regulator'' mass introduced into the
Hamiltonian of the 4+1--dimensional parent theory. In taking this
limit the fields of the parent theory are scaled to zero in such a
way that the Green's functions of 4--dimensional Euclidean theory
are obtained. We shall verify this firstly for a 1--flavour chiral
fermion and then for a 3--flavour example in which two of the
chiral fields combine to give a Dirac fermion. Further
generalization is then straightforward.

The basic recipe is to represent every Weyl doublet of the
4--dimensional theory by a 4--component Dirac spinor in the
4+1--dimensional theory. For one such field the parent Hamiltonian
(in the Schroedinger picture) is
$$
H=\int d^4x\ \hat{\bar\psi}(x)\ (\partial\!\!\! / +\Lambda
)\hat\psi (x)
$$
where $\hat\psi (x)$ is a 4--component spinor and $\hat{\bar\psi}(x)
=\psi (x)^\dagger\gamma_5$. The Dirac matrices $\gamma_\mu$ and
$\gamma_5$ are hermitian. The Schroedinger picture operators
satisfy the usual anticommutation rules,
$$
\{\hat\psi (x), \hat\psi (x')\}=0,\quad \{\hat\psi (x), \hat\psi
(x')^\dagger\}=\delta_4(x-x'),\quad
\{\hat\psi (x)^\dagger,\hat\psi (x')^\dagger\}=0
$$
They can be realized on a Fock space with vacuum state, $\vert
0\rangle$, annihilated by $\psi (x)$. The eigenstates of $H$ are
generated in the usual way by solving the 1--body problem,
$$
H(k)\ u(k,\lambda )=\omega (k)\ u(k,\lambda ),\quad H(k)\
v(k,\lambda )=-\omega (k)\ v(k,\lambda )
$$
where $H(k)=\gamma_5(ik\!\!\!/ +\Lambda )$ and $\omega (k)
=\sqrt{k^2+\Lambda^2}$ is positive. The eigenspinors, $u$ and $v$,
are orthonormal and complete since $H(k)$ is hermitian. We shall
write $u_\pm$ and $v_\pm$ to distinguish the cases $\Lambda
=\pm\vert\Lambda\vert$. Correspondingly, there are two independent
plane wave expansions
$$
\hat\psi (x) ={1\over\Omega}\ \sum_{k,\lambda}\left(b_\pm
(k,\lambda )u_\pm (k,\lambda)+d^\dagger_\pm (k,\lambda )v_\pm
(k,\lambda )\right) e^{ikx}
$$
where $\Omega$ is the volume of a 4--box. The operators $b_+$
and $d_+$ annihilate the Dirac vacuum, $\vert +\rangle$, defined by
$$
\vert +\rangle\ = \mathop{\Pi}\limits_{k,\lambda}\ \Omega^{-1/2}\
d_+(k,\lambda )\  \vert 0\rangle
$$
but, of course, they do not annihilate the other Dirac vacuum,
$$
\vert -\rangle\ = \mathop{\Pi}\limits_{k,\lambda}\ \Omega^{-1/2}\
d_-(k,\lambda )\ \vert 0\rangle
$$
The two Dirac vacua are normalized and their overlap is given by
\begin{eqnarray*}
\langle +\vert -\rangle & = &\mathop{\Pi}\limits_k\
{\rm det}_{\lambda\lambda '} (v^\dagger_+(k,\lambda ) v_-(k,\lambda
'))\\ & = & \mathop{\Pi}\limits_k\ (k^2/\omega^2)
\end{eqnarray*}
In order for this to be non--vanishing we must suppose that the
fermions are subject to antiperiodic boundary conditions.

Consider the Green's function defined by
\begin{eqnarray*}
G(x-x') & = & {\langle +\vert\hat\psi (x)\hat{\bar\psi} (x')\vert
-\rangle\over \langle +\vert -\rangle}\\
& = & {1\over\Omega^2}\ \Sigma\ u_+(k,\lambda )\
{\langle +\vert b_+(k,\lambda )b^\dagger_-(k',\lambda ')\vert
-\rangle\over\langle +\vert -\rangle}\ \bar u_-(k',\lambda ')
e^{ikx-ik'x'}\\
&=&{1\over\Omega}\ \sum_k\ \widetilde G(k)\ e^{ik(x-x')}
\end{eqnarray*}

To compute $\widetilde G(k)$ one needs the commutators
$\{
b_+,b^\dagger_-\}$, $\{b_+,d_-\}$, $\{d^\dagger_+,b^\dagger_-\}$
and $\{ d^\dagger_+,d_-\}$ which can all be expressed in terms of
the eigenspinors $u_\pm$ and $v_\pm$. The result is
$$
\widetilde G(k) ={1\over 2}\left(\omega +\gamma_5(ik\!\!\!/
+\Lambda )\right)\ {1\over ik\!\!\!/}
$$
Scaling with $1/\vert\Lambda\vert$ and taking the limit
$\Lambda\to\pm\infty$ gives
$${1\over\vert\Lambda\vert}\ \widetilde G(k)\to {1\over 2}
\left( 1\pm \gamma_5\right)\ {1\over ik\!\!\!/}
$$
i.e. the Euclidean Green's function for a chiral fermion. In
effect,
$$
\lim_{\Lambda\to\pm\infty}\ \left( \vert\Lambda\vert^{-1/2}\
\hat\psi (x)\right) ={1\over 2} (1\pm\gamma_5)\psi(x)
$$
where $\psi (x)$ represents a massless fermion in 4 dimensions.

This very elementary discussion can easily be extended to the
many--flavour case. With a view towards the Standard Model,
consider the 3--flavour neutrino--electron system, $\nu_L,e_L$ and
$e_R$. These fields are represented in the parent theory by three
4--component spinors $\hat\nu_L,\hat e_L$ and $\hat e_R$. The
Hamiltonian is
$$
H=\int d^4x\ \hat{\bar\psi}\left(\partial\!\!\!/ +\Lambda
T_c+\phi\cdot T\right)\hat\psi
$$
where
$$\hat\psi =\left(\matrix{ \hat\nu_L\cr \hat e_L\cr \hat e_R\cr}
\right)\quad {\rm and}\quad \Lambda T_c+\phi\cdot T=\left(
\matrix{ \Lambda &0&\phi_1\cr 0&\Lambda&\phi_2\cr
\phi^*_1&\phi^*_2&-\Lambda\cr}\right)
$$
\medskip

\noindent with $\phi_1,\phi_2$ representing the Higgs doublet: for
the following discussion we can replace these fields by vacuum
values,
$\langle\phi_1\rangle =0, \langle\phi_2\rangle =m$. Note the
$-\Lambda$ associated with $e_R$.

Eigenspinors of the 1--body Hamiltonian are to be computed and the
vacuum overlap is
$$
\langle +\vert -\rangle =\mathop{\Pi}\limits_k\
{\rm det}_{\sigma\sigma '} (v^\dagger_+(k,\sigma)v_-(k,\sigma '))
$$
where $\sigma$ and $\sigma '$ are 6--valued labels comprising
helicity and flavour. The eigenspinors can be computed
approximately by expanding in powers of $1/\Lambda$,
\begin{eqnarray*}
u_\pm (k,\sigma ) &=& u_\pm (\sigma )+{1\over 4\vert\Lambda\vert}
(1\mp\gamma_5T_c)\ V(k)u_\pm (\sigma )+\dots\\
v_\pm (k,\sigma ) &=& v_\pm (\sigma )+{1\over 4\vert\Lambda\vert}
(1\mp\gamma_5T_c)\ V(k)v_\pm (\sigma )+\dots
\end{eqnarray*}
where $V(k)=\gamma_5(ik\!\!\! /+\phi\cdot T)$ and $u_\pm(\sigma
)=v_\mp (\sigma )$ are constant spinors normalized such that
$$\sum_\sigma\ u_\pm (\sigma )\ u^\dagger_\pm (\sigma )={1\over
2}(1\pm\gamma_5T_c)
$$
The simple perturbative structure of the eigenspinors depends
crucially on the fact that $\gamma_5T_c$ anticommutes with $V(k)$.

The Green's function for this system is given by
\begin{eqnarray*}
\widetilde G(k) &=& {1\over\Omega }\ \sum_{\sigma ,\sigma
'}\ u_+(k,\sigma)\ {\langle +\vert b_+(k,\sigma
)b^\dagger_-(k,\sigma ')\vert -\rangle\over \langle +\vert
-\rangle }\ \bar u_-(k,\sigma ')\\
&=& \sum_{\sigma\sigma '}\ u_+(\sigma )\ (K^{-1})_{\sigma\sigma '}\
\bar u_-(\sigma ) +\dots
\end{eqnarray*}
to leading order in $\Lambda$, where the matrix $K$ is given by
\begin{eqnarray*}
K_{\sigma\sigma '} &=& {1\over \vert\Lambda\vert}\
u^\dagger_-(\sigma )\ V(k)\ u_+(\sigma )+\dots\\
&=& {1\over\vert\Lambda\vert}\ u^\dagger_-(\sigma )\
\gamma_5(ik\!\!\! / +\phi\cdot T)\ u_+(\sigma )+\dots
\end{eqnarray*}
Its inverse can be written
\begin{eqnarray*}
K^{-1}_{\sigma\sigma '} &=& \vert\Lambda\vert\
u^\dagger_+(\sigma )\ V(k)^{-1}\ u_-(\sigma ')+\dots\\
&=&\vert\Lambda\vert\ u^\dagger_+(\sigma )\
(ik\!\!\! / +\phi\cdot T)^{-1}\ \gamma_5\ u_-(\sigma ')+\dots
\end{eqnarray*}
so that, finally,
$$
\widetilde G(k)={\vert\Lambda\vert\over 2}\ (1+\gamma_5T_c)\
(ik\!\!\!/+\phi\cdot T)^{-1}+\dots\eqno(\ast)
$$
To interpret this we should rescale the fields to remove the
factor $\vert\Lambda\vert$ and then take the limit
$\Lambda\to\infty$,
$$
\lim_{\Lambda\to\infty}\ \left(\Lambda^{-1/2}\ \hat\psi
(x)\right)={1\over 2}\left( 1+\gamma_5T_c\right) \psi (x)
$$
The ``wrong chirality'' parts of the fields, $\gamma_5T_c=-1$, do
not contribute to the renormalized Green's functions in this
limit. The effective fields satisfy $\gamma_5T_c=1$, i.e. $\nu_L$
and $e_L$ have $\gamma_5=1$ while $e_R$ has $\gamma_5=-1$. If the
Higgs doublet acquires the usual vacuum value,
$\langle\phi_1\rangle =0$ and $\langle \phi_2\rangle =m$, then the
neutrino is massless and the electron is massive.

The structure ($\ast$) will be valid for any number of flavours
provided $\gamma_5T_c$ anticommutes with $V(k)$ i.e. if the
chirality matrix, $T_c$, anticommutes with the mass matrix,
$\phi\cdot T$.

\section{External gauge field and anomalies}

Having examined the free fermion Green's function we now include a
vector background by the minimal prescription
$\partial_\mu\to\nabla_\mu =\partial_\mu -i\ A_\mu (x)$, writing
$H(A)=H(0)+V$. The Dirac vacua are perturbed.
$$
\vert\pm\rangle\to\vert A\pm\rangle =\alpha_\pm (A)\left[
1-{1\over E_0-H_\pm (0)}\ \Pi_\pm (V-\Delta
E_\pm)\right]^{-1}\vert\pm\rangle
$$
where, again, the $\pm$ notation indicates the sign of $\Lambda$.
The operators $\Pi_\pm$ project out the unperturbed vacua,
$$
\Pi_\pm =1-\vert\pm\rangle\langle\pm\vert
$$
and the normalization factors $\alpha_\pm (A)$ are chosen to be
real and positive. The vacuum shifts, $\Delta E_\pm$, are
determined by the consistency conditions
$$
0=\langle\pm\vert (V-\Delta E_\pm )\vert A\pm\rangle
$$

The perturbed vacuum amplitude $\langle A+\vert A-\rangle$ depends
on the vector background and we must test its response to gauge
transformations. Acting on the Schroedinger picture fields these
transformations are realized by unitary operators, $U_\theta$,
$$
U^{-1}_\theta\ \hat\psi (x)\ U_\theta =e^{i\theta (x)}\ \hat\psi
(x)
$$
where $\theta (x)$ is an hermitian matrix acting in flavour space.
If the free Hamiltonian $H(0)$ is invariant under constant gauge
transformations then
$$
U^{-1}_\theta\ H(A)\ U_\theta =H(A^\theta )
$$
where $A^\theta_\mu =e^{-i\theta} (A_\mu +i\partial_\mu
)e^{i\theta}$. To keep the anomaly discussion as simple as
possible we discard the mass term, $\phi\cdot T$ and require only
that $\theta (x)$ commute with the chirality matrix, $T_c$.

If $A_\mu (x)$ is a weak, topologically trivial field then the
perturbed Dirac vacua must be non--degenerate. It follows that
$$
U_\theta\vert A\pm\rangle =\vert A^\theta\pm\rangle\ e^{i\Phi_\pm
(\theta ,A)}
$$
where the angles $\Phi_\pm$ are real. The group property,
$U_{\theta_1}\ U_{\theta_2}=U_{\theta_{12}}$ implies the
composition rule,
$$
\Phi (\theta_{12},A)=\Phi (\theta_1,A^{\theta_2})+\Phi (\theta_2,A)
$$
identically in $\theta_1,\theta_2$ and $A$. It is possible to
compute $\Phi_\pm (\theta ,A)$ by applying time independent
perturbation theory. To first order in $\theta$ this gives
$$
\Phi_\pm (\theta ,A)=\int_\Omega dx\
\langle\pm\vert\hat\psi^\dagger\ \theta\ \hat\psi\left[ 1-
{1\over E_0-H_\pm (0)}\ \Pi_{\pm}(V-\Delta
E_\pm)\right]^{-1}\vert\pm\rangle
$$
with the understanding that $\alpha_\pm (A)$ is real.

Since $U_\theta$ is unitary it follows that the vacuum amplitude
must satisfy
$$
\langle A^\theta +\vert A^\theta -\rangle =\langle A+\vert
A-\rangle\ e^{i\Phi_+(\theta ,A)-i\Phi_-(\theta ,A)}
$$
which implies an anomaly if $\Phi_+-\Phi_-\not= 2\pi n, n\in
Z\!\!\!Z$. We
compute this difference perturbatively and show that, in the limit
$\Lambda\to\infty$, it contains the usual (consistent) anomalies.
(The abelian anomaly in 2--dimensions was computed in an analogous
way in Ref.4.)

A complete discussion would require regularization of the theory
but we shall merely pick out the terms that contribute to the
anomaly since they will be expressible as convergent integrals. In
effect, we look for terms of second order in $A$ that contain the
tensor $\varepsilon_{\kappa\lambda\mu\nu}$. The second order part
of $\Phi_+$ is given by
$$\Phi^{(2)}_+=\sum_{n,m>0}\int_\Omega dx\ \langle
+\vert\hat\psi^\dagger\ \theta\ \hat\psi\vert n\rangle\ {1\over
E_0-E_n}\ \langle n\vert V\vert m\rangle\ {1\over E_0-E_m}\
\langle m\vert V\vert +\rangle
$$
where the sums are restricted to 2--particle intermediate states,
$$\sum_n\vert n\rangle\langle n\vert ={1\over\Omega^2}\ \Sigma\
\vert k\sigma ,\overline{k'\sigma '}\rangle\langle k\sigma ,
\overline{k'\sigma '}\vert
$$
where
$$
\vert k\sigma ,\overline{k'\sigma '}\rangle =b^\dagger_+(k,\sigma
)\ d^\dagger_+(k',\sigma ')\vert +\rangle
$$
In the infinite volume limit this reduces to
\begin{eqnarray*}
\Phi^{(2)}_+ & = & \int dx\int\left({dp_1\over 2\pi}\ {dp_2\over
2\pi}\ {dk\over 2\pi}\right)^4
\Biggl[{{\rm tr} (\theta (x)U(k)i\tilde A\!\!\!/
(p_1)\gamma_5U(k-p_1)i\tilde A\!\!\!/ (p_2)\gamma_5
V(k-p_1-p_2))\over (\omega (k)+\omega (k-p_1-p_2))(\omega
(k-p_1)+\omega (k-p_1-p_2))} -\Biggr.\\
&& \\
&&\Biggl.-{{\rm tr} (\theta (x)U(k)i\tilde A\!\!\!/
(p_1)\gamma_5V(k-p_1)i\tilde A\!\!\!/ (p_2)\gamma_5
V(k-p_1-p_2))\over (\omega (k)+\omega (k-p_1-p_2))(\omega
(k)+\omega (k-p_1))}\Biggr] e^{i(p_1+p_2)x}
\end{eqnarray*}
where $\omega (k)=\sqrt{k^2+\Lambda^2}$ and
$$
U(k)={\omega_k+\gamma_5(ik\!\!\! /+\Lambda )\over 2\omega_k} =
1-V(k)
$$
To obtain $\Phi^{(2)}_-$, reverse the sign of $\Lambda$.
Evalutating the Dirac trace and discarding terms that do not
contain the antisymmetric tensor one finds
\begin{eqnarray*}
\Phi^{(2)}_\pm &=&\pm{\vert\Lambda\vert\over 2}\int dx\int\left(
{dp_1\over 2\pi}\ {dp_2\over 2\pi}\ {dk\over 2\pi}\right)^4
{1\over\omega (k)\omega (k-p_1)\omega (k-p_1-p_2)}\
{1\over \omega (k)+\omega (k-p_1-p_2)}\cdot\\
&& \\
&&\cdot\left({1\over \omega (k-p_1)+\omega (k-p_2)}+
{1\over\omega (k)+\omega (k-p_1)}\right)
\varepsilon_{\kappa\lambda\mu\nu}\ p_{1\kappa}p_{2\lambda}\ {\rm
tr}\left(\theta (x)\tilde A_\mu (p_1) \tilde A_\nu(p_2)\right)
\end{eqnarray*}
where the trace here is restricted to flavour space. The integral
over $k$ converges and can be estimated at large
$\vert\Lambda\vert$ by setting $p_1=p_2=0$ to obtain
$$
\int\left({dk\over 2\pi}\right)^4\ {1\over\omega (k)^5}={1\over
12\pi^2\vert\Lambda\vert}
$$
Hence, for $\vert\Lambda\vert\to\infty$,
$$
\Phi^{(2)}_\pm (\theta ,A)\to\pm{1\over 48\pi^2}\int dx\
\varepsilon_{\kappa\mu\lambda\nu}\ {\rm tr}\left(
\theta (x)\ \partial_\kappa\ A_\mu\ \partial_\lambda\ A_\nu\right)
$$
and we see that the difference, $\Phi_+-\Phi_-$, indeed contains
the usual consistent anomaly.

To construct anomaly--free models it is necessary to combine
chiral multiplets in a suitably way. In effect this means choose a
chirality matrix, $T_c$, that commutes with $\theta (x)$ and
satisfies
$$
\varepsilon_{\kappa\mu\lambda\nu}\ {\rm tr}\left( T_c\ \theta (x)\
\partial_\kappa\ A_\mu\ \partial_\lambda\ A_\nu\right) =0
$$
For example for one family of quarks
and leptons in the standard $SU(2)_L\times U(1)_Y$ model choose
$T_c^{lep} =diag (1, 1, -1)$ and $T_c^{quark}=diag(1, 1, -1, -1)$
for each colour.

\end{document}